\documentstyle[titlepage,12pt]{article}
\begin{document}
\renewcommand{\theequation}{\thesection.\arabic{equation}}
\title{Circular Strings and Multi-Strings in de Sitter and
Anti de Sitter Spacetimes\thanks{Research report to appear in
{\bf "New developments in string
gravity and physics at the Planck energy scale"}
Edited by N. S\'{a}nchez (World
Scientific, Singapore, 1995)}}
\author{H.J. de Vega\\
Laboratoire de Physique
Th\'{e}orique et Hautes Energies.\\
Laboratoire Associ\'{e} au CNRS
UA280, Universit\'{e} de Paris VI et VII,\\
Tour 16,
1er \'{e}tage, 4, Place Jussieu,
75252 Paris, France.\\
and \\
A.L. Larsen,$\;$ N. S\'{a}nchez\\
Observatoire de Paris,
DEMIRM.
Laboratoire Associ\'{e} au\\
CNRS UA 336,
Observatoire de Paris et
\'{E}cole Normale Sup\'{e}rieure.\\
61, Avenue
de l'Observatoire, 75014 Paris, France.}
\maketitle
\begin{abstract}
The exact general solution of circular strings in $2+1$
dimensional de Sitter spacetime is described completely in terms of elliptic
functions. The novel feature here is that one single world-sheet generically
describes {\it infinitely many} (different and independent) strings. This has
no analogue in flat spacetime. The circular strings are either oscillating
("stable") or indefinitely expanding ("unstable").
We then compute the {\it exact} equation of
state of circular strings in the $2+1$ dimensional de Sitter (dS) and anti
de Sitter (AdS) spacetimes, and analyze its properties for the different
(oscillating, contracting and expanding) strings. We finally perform a
semi-classical quantization of the oscillating circular strings.
The string mass is $m=\sqrt{C}/(\pi H\alpha'),\;C$
being the Casimir operator, $C=-L_{\mu\nu}L^{\mu\nu},$ of the $O(3,1)$-dS
[$O(2,2)$-AdS] group, and $H$ is the Hubble constant.
We find the mass formula
$\alpha'm^2_{\mbox{dS}}\approx 4n-5H^2\alpha'n^2,\;(n\in N_0),$ and a
{\it finite} number of states $N_{\mbox{dS}}\approx 0.34/(H^2\alpha')$ in
de Sitter spacetime; $m^2_{\mbox{AdS}}\approx H^2n^2$ (large $n\in N$) and
$N_{\mbox{AdS}}=\infty$ in anti de Sitter spacetime. The level spacing
grows with $n$ in AdS spacetime, while is approximately constant (although
smaller than in Minkowski spacetime and slightly decreasing) in dS spacetime.
\end{abstract}
\section{Introduction}
\setcounter{equation}{0}
The study of string dynamics in curved spacetimes reveals new insights and
new physical phenomena with respect to string propagation in flat spacetime
(and with respect to quantum fields in curved spacetimes)
\cite{veg1,veg2,veg3}.
The results of
this programme are relevant both for fundamental (quantum) strings and for
cosmic strings, which behave essentially in a classical way.

Among the cosmological backgrounds of interest, de Sitter spacetime occupies
a special place. It is on one hand relevant for inflation, and on the
other hand, string propagation turns out to be specially interesting there.
String-instability, in the sense that the string proper length
grows indefinitely (proportional to the expansion factor of the universe) is
particularly present in de Sitter spacetime [1-8].
Moreover, a novel feature for strings in curved spacetimes
was first found in de Sitter spacetime: Exact {\it multi-string} solutions
\cite{mic2,mic1}. That is, one single world-sheet generically describes
two strings \cite{mic1}, several strings \cite{mic2} or even
{\it infinitely many} (different and independent) strings \cite{all1}

Circular strings are specially suited for detailed
investigation. Since the string equations of motion become separable, one has
to deal with non-linear ordinary differential equations instead of
non-linear partial differential equations. In order to obtain
generic non-circular
string solutions the full power of the inverse scattering method is
needed in de Sitter spacetime \cite{mic2}.

Here we will present some more recent results \cite{all1,all2} on exact
circular string solutions
in de Sitter spacetime and, for comparison, in anti de Sitter spacetime too.
That is, we consider
the circular string solutions of the equations of motion and
constraints:
\begin{equation}
\ddot{x}^\mu-x''^\mu+\Gamma^\mu_{\rho\sigma}(\dot{x}^\rho\dot{x}^\sigma-
x'^\rho x'^\sigma)=0,
\end{equation}
\begin{equation}
g_{\mu\nu}\dot{x}^\mu x'^\nu=g_{\mu\nu}(\dot{x}^\mu\dot{x}^\nu+x'^\mu x'^\nu)
=0,
\end{equation}
where dot and prime stand for derivative with respect to the
world-sheet coordinates $\tau$ and
$\sigma,$ respectively and $\Gamma^\mu_{\rho\sigma}$ are the Christoffel
symbols
with respect to the metric $g_{\mu\nu}.$
The exact general solution of circular strings in $2+1$ dimensional
de Sitter spacetime
is described closely and completely in terms of elliptic
functions (Section 2).

By computing the string energy and pressure, we obtain the corresponding
equations of state, providing the physical interpretation of the
solutions in a cosmological context.
We analyze
the equations of state
for the different (oscillating, contracting and expanding)
strings. The string equation of state turns out to have
the perfect fluid form
$P=(\gamma-1)E,$ with the instantaneous coefficient $\gamma$
depending on an elliptic modulus (Section 3).

Cosmological backgrounds like de Sitter and anti de Sitter spacetimes
are not Ricci flat and hence they are not
string vacua even at first order in $\alpha'$. Strings are there
non-critical and quantization will presumably lead to features like
ghost states. No definite answer is available by now to such
conformal anomaly effects.

It is important in this context to investigate the quantum
aspects in the semi-classical regime,
where anomaly effects are practically irrelevant.
Semi-classical, in this context, means the regime in which $H^2\alpha'<<1,$
where $H$ is the Hubble constant.
We semi-classically quantize the time-periodic string
solutions in de Sitter and anti de Sitter spacetimes. Time-periodic string
solutions here include all the circular string solutions in anti
de Sitter spacetime, as well as the oscillating string solutions in de Sitter
spacetime (Section 4).
\section{Circular Strings in de Sitter Spacetime, Multi-Strings}
\setcounter{equation}{0}
Recently, several progresses in the understanding of string propagation in
de Sitter spacetime have been obtained [6-11].
The classical string
equations of motion (plus the string constraints) were shown to be
integrable in D-dimensional
de Sitter spacetime \cite{san1,san2}. They are equivalent to a non-linear
sigma model on the Grassmannian $SO(D,1)/O(D)$ with periodic boundary
conditions (for a closed string). In addition, the string constraints
imply a zero world-sheet energy-momentum tensor, and these constraints
are compatible
with the integrability. Moreover, the exact string dynamics in de Sitter
spacetime is equivalent to a generalized sh-Gordon model with a potential
unbounded from below \cite{san2}. The sh-Gordon function $\alpha(\sigma,\tau)$
has here a clear physical meaning: $H^{-1}\exp[\alpha(\sigma,\tau)/2]$
determines the proper size of the string ($H$ is the Hubble constant). In
$2+1$ dimensions, the string dynamics is exactly described by the
standard sh-Gordon equation.

More recently, a novel feature for strings in de Sitter spacetime
was found: Exact multi-string solutions
\cite{mic2,mic1}. Exact circular string solutions were found describing two
different strings \cite{mic1}. One
string is stable (the proper size is bounded),
and the other one is unstable (the proper size blows up) for large
de Sitter radius. Soliton methods, (the so-called "dressing method" in
soliton theory) were implemented using the linear problem (Lax pair) of
this system, in order to construct systematically exact string solutions
\cite{mic2}. The one-soliton string solution constructed in this way,
generically describe five different and independent strings: one stable
string and four unstable strings. These solutions (even the stable string)
do not oscillate in time.

In this section, we go further in the investigation of exact string solutions
in de Sitter spacetime. We find exact string solutions describing
{\it infinitely many} different and independent strings. The novel feature
here is that we have one single world-sheet but multiple (infinitely many)
strings. The world-sheet time $\tau$ turns out to be an infinite-valued
function of the target space time $X^0$ (which can be the hyperboloid time
$q^0$, the comoving time $T$ or the static coordinate time $t$).
Each branch of $\tau$ as a
function of $q^0$ corresponds to a different string. In flat spacetime,
multiple string solutions are necessarily described my multiple
world-sheets. Here, a single world-sheet describes infinitely many different
and simultaneous strings as a consequence of the coupling to the
spacetime geometry. These strings do not interact among themselves; all the
interaction is with the curved spacetime.
\vskip 6pt
\hspace*{-6mm}We apply the circular string {\it Ansatz}:
\begin{equation}
t=t(\tau),\;\;\;\phi=\sigma,\;\;\;r=r(\tau),
\end{equation}
in $2+1$ dimensional de Sitter spacetime, particularly convenient in terms
of the static de Sitter coordinates $(t,r,\phi)$ (we also describe the
solutions in the hyperboloid and comoving
parametrizations). The string equations
of motion and constraints, Eqs.(1.1)-(1.2),
can be solved directly and completely in terms of
elliptic functions. They reduce to two decoupled first order differential
equations for
the time component $t(\tau)$ and the string radius $r(\tau)$:
\begin{equation}
\dot{t}=\frac{\sqrt{b}\alpha'}{1-H^2r^2}
\end{equation}
\begin{equation}
\dot{r}^2+V(r^2)=b\alpha'^2;\;\;\;\;V(r^2)=r^2(1-H^2r^2)
\end{equation}
Notice that we are using the notation of Ref.\cite{all2}, which is slightly
different from the notation of Ref.\cite{all1}.
The $\dot{r}$-equation is solved by:
\begin{equation}
H^2r^2(\tau)=\wp(\tau-\tau_o)+1/3,
\end{equation}
where
$\wp$ is the Weierstrass elliptic function \cite{abr} with discriminant
$\Delta=16b^2H^4\alpha'^4(1-4bH^2\alpha'^2)$,
and $b$ and $\tau_o$ are integration constants
($\tau_o\;$ is generally complex and must be chosen such that $r(\tau)$ is
real for real $\tau$). The solutions depend on one constant parameter $b$
(for fixed $H,\alpha'$) related to the string energy, see Section 3,
and fall into three classes, depending on
whether:
\begin{eqnarray}
bH^2\alpha'^2<1/4\;(\Delta>0),\nonumber\\
bH^2\alpha'^2=1/4\;(\Delta=0),\nonumber\\
bH^2\alpha'^2>1/4\;(\Delta<0).\nonumber
\end{eqnarray}
As can be seen in the diagram $(r^2,V(r^2))$, Fig.1., in which the full string
dynamics takes place, these cases correspond to either oscillatory motion or
infinite (unbounded) motion.

The proper string size $S$ of the circular strings are given for all $r$ by:
\begin{equation}
S(\tau)=r(\tau),
\end{equation}
as can be seen from the induced line element on the string world-sheet:
\begin{equation}
ds^2=r^2(\tau)(-d\tau^2+d\sigma^2).
\end{equation}
\vskip 6pt
\hspace*{-6mm}In the $bH^2\alpha'^2=1/4$ case, the Weierstrass elliptic
function
degenerates into a hyperbolic function:
\begin{equation}
H^2r^2(\tau)=\frac{1}{2}[1+\sinh^{-2}(\frac{\tau-\tau_o}{\sqrt{2}})].
\end{equation}
Two real independent solutions appear for the choices $\tau_o=i\pi/2$ and
$\tau_o=0$, respectively:
\begin{equation}
H^2r^2_\pm(\tau)=\frac{1}{2}[\tanh(\frac{\tau}{\sqrt{2}})]^{\pm 2}.
\end{equation}
(They were first found in Ref.\cite{mic1}).
We have also the solution $H^2r^2_0=1/2$, corresponding to a stable string with
constant proper size $S_0=1/(\sqrt{2}H)$ (i.e., sh-Gordon function
$\alpha=0$). This solution was found in Ref.\cite{mic1} and we shall not
discuss
it here.

The solution $r_-$ describes two different strings, I and II, as it can be
seen from the hyperboloid time $q^0_-(\tau),$  Fig.2.
Here $\tau$ is a two-valued function of $q^0_-$: String I
corresponds to $-\infty<\tau<0$ and string II to $0<\tau<\infty$. The
proper size $S_-$ for both strings is given by
Eq.(2.5), using Eq.(2.8).
For $q^0_-\rightarrow\infty$, string I is unstable, while
string II is stable. The proper size $S_-(q^0_-\rightarrow\infty)$
blows up for string I (for which
$q^0_-\rightarrow\infty$ corresponds to $\tau\rightarrow 0_-$), while
tends to a constant value for string II (for which $q^0_-\rightarrow\infty$
corresponds to $\tau\rightarrow\infty$). String I starts with minimal size
$S_-=1/(\sqrt{2}H)$ at $\tau=-\infty$ and blows up
at $\tau=0$. String II starts with infinite size at $\tau=0$ but approaches
$S_-=1/(\sqrt{2}H)$ for $\tau\rightarrow\infty$.

The solution $r_+$ of this $bH^2\alpha'^2=1/4$ case
describes only one stable string. The hyperboloid time $q^0_-(\tau)$,
is a monotonically increasing function of $\tau$
and the corresponding string has
bounded proper size $S_+.$ The string
starts with $S_+=1/(\sqrt{2}H)$, at $q^0_+=-\infty$,
it contracts until it collapses ($S_+=0$), then it
expands until it reaches the original size for $q^0_+=
\infty$. For $bH^2\alpha'^2=1/4$ the evolution is
always
non-oscillatory. Even the stable string does not oscillate in time.
\vskip 6pt
\hspace*{-6mm}For $bH^2\alpha'^2<1/4$ there exist two real independent
solutions
for the choices $\tau_o=0$ and $\tau_o=\omega'$, where
$\omega'$ is the
imaginary semi-period of the Weierstrass function:
\begin{equation}
H^2r^2_-(\tau)=\wp(\tau)+1/3,
\end{equation}
\begin{equation}
H^2r^2_+(\tau)=\wp(\tau+\omega')+1/3.
\end{equation}
$r_-$ and $r_+$ are oscillating solutions as functions of $\tau$. The
solution $r_-$ describes infinitely many strings; $r_-$ has infinitely
many branches $[0,2\omega],\;[2\omega,4\omega],\;...$, each of which
corresponds
to a different string ($\omega$ is the real semi-period, of the
Weierstrass function). This can be seen from the hyperboloid time
$q^0_-(\tau)$,
Fig.3:
The world-sheet time $\tau$ is an infinite-valued function of
$q^0_-$. The hyperboloid time $q^0_-$ blows up at the boundaries of the
branches $\tau=\pm 2N\omega\;$ ($N$ being an integer):
\begin{equation}
\mid q^0_-(\tau)\mid\sim\frac{1}{\mid 2N\omega-\tau\mid}.
\end{equation}
Further insight is obtained by considering the comoving time $T_-$ and the
static coordinate time $t_-$. Closed expressions for them are given in
Ref.\cite{all1}, in terms of Weierstrass $\zeta$ and
$\sigma$-functions, and also rewritten in terms of elliptic theta-functions.
The cosmic time $T_-$ is singular at $\tau=0,\;\tau=x/\mu,\;\tau=2\omega$
and similarly in the other branches. ($x,\;\mu$ are two real constants. $x$
is expressed as an incomplete elliptic integral of first kind while
$\mu=\sqrt{(1+\sqrt{1-4bH^2\alpha'^2})/2}\;$).
The static coordinate time $t_-$, on the
other hand, is regular at the boundaries of the branches, but is singular
at $\mu\tau=2KN\pm x$:
\begin{equation}
t_-(\tau)\sim\frac{1}{2\pi}\log\mid\mu\tau-2KN\mp x\mid,
\end{equation}
where $K$ is a complete elliptic integral of the first kind. It must be
noticed that although $r_-$ is periodic in $\tau$, the comoving time is not,
i.e. $T_-(\tau)\neq T_-(\tau+2\omega).$ This implies that the
infinitely many strings are different. The difference in proper size of the
$n$'th and the $(n+1)$'th string for a given comoving time $T_-$ is given by:
\begin{equation}
\Delta S_-=\frac{\pi}{H} \frac{\theta_1'}{\theta_1}
(\frac{\pi x}{2K})e^{HT_-}e^{-n\pi\frac{\theta_1'}{\theta_1}
(\frac{\pi x}{2K})}.
\end{equation}
However,
{\it all} the strings are
of the same type: {\it unstable}. For instance, in the
branch $\tau\in
[0,2\omega]$, the string starts at $\tau=0\;(q^0_-=-\infty)$ with infinite
size, then contracts to the minimal size
$HS_-=\sqrt{(1+\sqrt{1-4bH^2\alpha'^2})/2}$
and eventually expands towards infinite size at $\tau=2\omega\;(q^0_-=
\infty)$. These solutions never collapse to $r=0.$

For the solution $r_+$ of the $bH^2\alpha'^2<1/4$ case, the string
dynamics takes place inside the
horizon. $r_+$, being a regularly oscillating function of $\tau$,
is then also a regularly oscillating function of the physical times
$q^0_+,\;T_+$ and $t_+$. The static coordinate time $t_+$, from which one
easily deduces $q^0_+$ and $T_+$, is expressed in terms of theta-functions
\cite{all1}.
The solution $r_+$ describes one stable string oscillating between its
minimal size $S_+=0$ (collapse) and its maximal size $HS_+=\sqrt{(1-
\sqrt{1-4bH^2\alpha'^2})/2}$.
It must be noticed that the string oscillations here do
not follow a pure harmonic motion as in flat Minkowski spacetime, but they
are precise superpositions of all frequencies $(2n-1)\Omega,\;(\Omega
=\pi\mu/(2K),\;n=1,2,...,\infty)$ \cite{all1}; here
the non-linearity of the string equations of motion
fixes the relation between the mode coefficients, and the basic frequency
$\Omega$ depends on the string energy via $b.$
\vskip 6pt
\hspace*{-6mm}For $bH^2\alpha'^2>1/4$
two real independent solutions are obtained for
$\tau_o=0$ and $\tau_o=\omega_2'$, where $\omega_2'$ is the imaginary
semi-period of the Weierstrass function:
\begin{equation}
H^2r^2_-(\tau)=\wp(\tau)+1/3,
\end{equation}
\begin{equation}
H^2r^2_+(\tau)=\wp(\tau+\omega_2')+1/3.
\end{equation}
In this case $r_-$ again describes infinitely many strings, all of them
are unstable. The difference with the $bH^2\alpha'^2<1/4$ case,
is that here the strings
have a collapse during their evolution. For instance, in the branch
$\tau\in[0,2\omega_2]$, where $\omega_2$ is the real semi-period of the
Weierstrass function, the string starts with infinite size at
$\tau=0\;(q^0_-=-\infty)$, it then contracts until it collapses to a point
and then it expands towards infinite size again at $\tau=2\omega_2\;
(q^0_-=\infty).$ In contrast to the $bH^2\alpha'^2<1/4$ case,
the solution $r_+$ is
here just a time translated version of $r_-$:
\begin{equation}
r^2_+(\tau)=r^2_-(\tau+\omega_2),
\end{equation}
and describes therefore essentially the same features as the solution $r_-$.

A summary of the main features and conclusions of the results of this
section is presented
in Table 1. Further details can be found in Ref.\cite{all1}.
\section{Physical Interpretation, Energy, Pressure}
\setcounter{equation}{0}
In this section we discuss the physical properties (energy, pressure) of
a gas of circular strings in de Sitter and anti de Sitter
spacetimes. In static coordinates, we thus consider the spacetimes:
\begin{equation}
ds^2=-a(r)dt^2+\frac{dr^2}{a(r)}+r^2 d\phi^2.
\end{equation}
For simplicity we consider the string dynamics in a 2+1 dimensional spacetime.
All our solutions can however be embedded in a higher dimensional spacetime,
where they will describe plane circular strings. We are interested in the
three cases:
\begin{eqnarray}
a(r) &=& 1 \quad{\rm for~Minkowski~ spacetime},\nonumber \\
a(r)&=&1-H^2r^2 \quad{\rm for~de~ Sitter~ spacetime},\nonumber \\
a(r)&=&1+H^2r^2 \quad{\rm for~anti~ de~ Sitter ~spacetime}.\nonumber
\end{eqnarray}
The equations describing the evolution of circular strings are:
\begin{equation}
\dot{t}=\frac{\sqrt{b}\alpha'}{a(r)}
\end{equation}
\begin{equation}
\dot{r}^2+V(r)=b\alpha'^2;\;\;\;\;V(r)=r^2a(r)
\end{equation}
and generalize Eqs.(2.2)-(2.3).

Properties like
energy and pressure of the strings are more conveniently discussed in
comoving (cosmological) coordinates:
\begin{equation}
ds^2=-(dT)^2+a^2(T)\;\frac{dR^2+R^2 \,d\phi^2}{(1+\frac{k}{4}R^2)^2},
\end{equation}
including as special cases Minkowski, de
Sitter and anti de Sitter
spacetimes:
\begin{eqnarray}
a(T) &=& 1,\;\;k=0 \quad{\rm for~Minkowski~ spacetime},\nonumber \\
a(T)&=&e^{HT},\;\;k=0 \quad{\rm for~de~ Sitter~ spacetime},\nonumber \\
a(T)&=&\cos HT,\;\;k=-H^2
\quad{\rm for~anti~ de~ Sitter ~spacetime}.\nonumber
\end{eqnarray}
The spacetime energy-momentum tensor is given in $2+1$ dimensions by:
\begin{equation}
\sqrt{-g}\,T^{\mu\nu}=\frac{1}{2\pi\alpha'}\int d\tau d\sigma\;(\dot{X}^\mu
\dot{X}^\nu-X'^\mu X'^\nu)\;\delta^{(3)}(X-X(\tau,\sigma)).
\end{equation}
After integration over a spatial volume
that completely encloses the string \cite{cos}, the
energy-momentum tensor for a circular string takes the form of a fluid:
\begin{equation}
T^\mu_\nu=\mbox{diag.}(-E,\;P,\;P),
\end{equation}
where, in the comoving coordinates (3.4):
\begin{equation}
E(X)=\frac{1}{\alpha'}\,\dot{T},
\end{equation}
\begin{equation}
P(X)=\frac{1}{2\alpha'}\,\frac{a^2(T)}{(1+\frac{k}{4}R^2)^2}\;
\frac{\dot{R}^2-R^2}{\dot{T}},
\end{equation}
represent the string energy and pressure, respectively.

We have already described all the circular string solutions in de Sitter
spacetime, Section 2.
The circular string solutions in anti de Sitter spacetime were obtained in
Ref.\cite{ads}. The general solution is:
\begin{equation}
Hr(\tau)=\frac{\bar{k}}{\sqrt{1-2\bar{k}^2}}\mbox{cn}
[\frac{\tau}{\sqrt{1-2\bar{k}^2}},\;\bar{k}],
\end{equation}
where:
\begin{equation}
\bar{k}^2=\frac{-1+\sqrt{1+4bH^2\alpha'^2}}{\sqrt{1+4bH^2\alpha'^2}}.
\end{equation}
For arbitrary values of the elliptic modulus $\bar{k}\in[0,\;\sqrt{1/2}\;[\;$,
the solution describes one oscillating string.

Let us now return to the energy-momentum tensor for circular strings.
The circular string solutions in de Sitter and anti de Sitter
spacetimes depend on the elliptic modulus
$k\equiv\sqrt{1-\mu^2}\;/\mu$ and $\bar{k}$,
respectively, as we have seen.
{}From these exact solutions we find the corresponding energy-momentum
tensors. They turn out to have the perfect fluid form with an equation of
state:
\begin{equation}
P=(\gamma-1)E,
\end{equation}
where $\gamma$ in general is time-dependent and depends on the elliptic
modulus ($k$ or $\bar{k}$)
as well.
We have analyzed the equation of state, Eq.(3.11), for all circular string
solutions in de Sitter and anti de Sitter spacetimes, Ref.\cite{all2}.

In de Sitter
spacetime, for strings expanding from zero radius towards infinity, the
equation of state changes continuosly from the ultra-relativistic matter-type
$P=+E/2,$ (in $2+1$ dimensions) when $r\approx 0,$ to the unstable
string-type \cite{ven1,ven2}, $P=-E/2,$ when $r\rightarrow\infty,$ see Fig.4.

On the other hand, for an
oscillating stable string in de Sitter spacetime, $\gamma$
oscillates between $\gamma(r=0)=3/2$ and
$\gamma(r=r_{\mbox{max}})=1/2+k^2/(1+k^2),$ where $k\in[0,\;1].$
Averaging over one oscillation period, the pressure {\it vanishes} \cite{all2}.
That is, these stable string solutions actually describe (in average)
{\it cold matter}. See also Fig.5.

In anti de Sitter spacetime, only
the oscillating (stable) circular string solutions, Eq.(3.9), exist.
We find that $\gamma$ oscillates between $\gamma(r=0)=3/2$ and
$\gamma(r=r_{\mbox{max}})=1/2,$ i.e. the equation of state "oscillates"
between $P=+E/2$ and $P=-E/2.$ This is similar to the situation in flat
Minkowski spacetime. When averaging over an oscillation period in anti
de Sitter spacetime, we find that $\gamma$
takes values from $1$ to $1+1/\pi^2$ for the allowed range of the elliptic
modulus. That is, the average pressure over one oscillation period is
always {\it positive} in anti de Sitter spacetime.

In conclusion, positive pressure characterizes the regime in which the string
radius is small relative to the string maximal size, while negative
pressure is characteristic for the regime in which the string radius is large.
In Minkowski spacetime, the two regimes are of equal "size", in the sense that
the average pressure is identically zero. The influence of the spacetime
curvature is among other effects, to modify the relative "size" of the two
regimes.

A summary of these features are presented in Table 2. More details are given
in Ref.\cite{all2}.
\section{Semi-Classical Quantization}
\setcounter{equation}{0}
In this section we perform a semi-classical quantization of the circular
string configurations discussed in the previous sections. We use an approach
developed in field theory by Dashen et. al. \cite{dhn,hm}, based
on the stationary phase approximation of
the partition function. The method can be only used for time-periodic
solutions of
the classical equations of motion. In our string problem, these
solutions however, include all the circular
string solutions in Minkowski and in anti de Sitter spacetimes, as well as the
oscillating circular strings ($bH^2\alpha'^2\leq 1/4,$ c.f. Section 2)
in de Sitter spacetime.

The result of the stationary phase integration is expressed in terms of
the function:
\begin{equation}
W(m)\equiv S_{\mbox{cl}}(T(m))+m\;T(m),
\end{equation}
where $S_{\mbox{cl}}$ is the action of the classical solution, $m$ is the
mass and
the period $T(m)$ is implicitly given by:
\begin{equation}
\frac{dS_{\mbox{cl}}}{dT}=-m.
\end{equation}
In string theory we must choose $T$ to be the period in a physical
time variable. For example, when a light cone gauge exists, $T$ is the
period in $X^0=\alpha' p\tau.$
The bound state quantization condition then becomes \cite{dhn,hm}:
\begin{equation}
W(m)=2\pi n,\quad n \in N_{0}
\end{equation}
for $n$ `large'. The method has been successfully used in many cases from
quantum mechanics to quantum field theory. For integrable
field theories the semi-classical quantization happens in fact,
to be exact. It must be noticed that string
theory in de Sitter spacetime is exactly integrable \cite{san1,san2}.
\vskip 6pt
\hspace*{-6mm}The actual computation
of $W(m)$ is straightforward but tedious; the details
are given in Ref.\cite{all2}.

For completeness we first considered flat Minkowski spacetime. In
that case, Eq.(4.3) reproduces the exact mass
spectrum, except for the intercept. That is, the mass spectrum becomes
$\alpha'm^2=4\;n,\;\;\;n\in N_0.$

We find for de Sitter (anti de Sitter) spacetime that the mass is
exactly proportional to  the square-root of the Casimir operator
$C=-L_{\mu\nu}L^{\mu\nu}$ of the $O(3,1)$-de Sitter
[$O(2,2)$-anti de Sitter] group:
\begin{equation}
m = {\sqrt{C} \over {\pi  H \alpha'}}.
\end{equation}
In Fig.6 we give parametric plots of
$H^2\alpha'W$ as a function of $H^2m^2\alpha'^2$ for $k\in[0,\;1]$
for  de Sitter spacetime and for
${\bar k}\in[0,\;1/\sqrt{2}\;[\;$  for anti
de Sitter spacetime, respectively.
We find for de Sitter spacetime:
\begin{equation}
\alpha'm^2_{\mbox{dS}}\approx 4\;n-5H^2\alpha'\;n^2,\quad n\in N_{0}
\end{equation}
This is different from the mass spectrum in Minkowski spacetime.
The level spacing is however still approximately
constant, but the levels are less separated than in Minkowski spacetime.
Notice in particular that the level spacing slightly decreases for larger
and larger $n.$ In de Sitter spacetime
there is only a {\it finite} number of levels as can be seen from
Fig.1.  The number of
quantized circular string states can be estimated to be:
\begin{equation}\label{numm}
N_q \approx\frac{0.34}{H^2\alpha'}.
\end{equation}
It is interesting to compare this result with the number of particle
states obtained using canonical quantization \cite{veg1}. One finds in
this way a  maximum number of states:
\begin{equation}
N_{\mbox{max}}\approx \frac{0.15}{H^2\alpha'},
\end{equation}
which is of the same order as the semi-classical value Eq.(4.6). It must be
noticed that in de Sitter spacetime, these states can {\it decay} quantum
mechanically due to the possibility of quantum mechanical tunneling through
the potential barrier, see Fig.1. Semi-classically, the decay probability
is however highly suppressed for $H^2\alpha'<<1$ and for any value of the
elliptic modulus $k,$ except near $k=1$ where the barrier disappears,
and for which the tunneling probability is close to unity, see Ref.\cite{all2}.

In anti de Sitter spacetime arbitrary high mass states exist.
The quantization of the high mass states yields (see Ref.\cite{all2}):
\begin{equation}
\alpha' m^2_{\mbox{AdS}}\approx H^2\alpha'\;n^2.
\end{equation}
Thus the (mass)$^2$ grows like $n^2$ and the level spacing grows
proportionally to $n.$ This is a completely different behaviour as compared
to Minkowski spacetime where the level spacing is constant.
A similar result was
recently found, using canonical quantization of generic strings
in anti de Sitter spacetime \cite{ads,all}. The
physical consequences, especially
the non-existence of a critical string temperature (Hagedorn
temperature), of this kind of behaviour
is discussed in detail in Ref.\cite{all}.

For both de Sitter and anti de Sitter spacetimes we find thus a
strong qualitative
agreement between the results obtained using canonical quantization,
based on generic string solutions (string perturbation approach), and the
results obtained using the semi-classical approach, based on oscillating
circular string configurations.

A summary of these features are presented in Table 3. More details are given
in Ref.\cite{all2}.
\section{Conclusion}
\setcounter{equation}{0}
We have found the exact general evolution of circular strings in $2+1$
dimensional de Sitter spacetime. We have expressed it closely and completely
in terms of elliptic functions, see Ref.\cite{all1} for the details.
The solution generically describes
infinitely many (different and independent) strings, and depends on one
constant parameter $b$ (for fixed $H,\alpha'$) related to the string energy.
We have computed {\it exactly} the equation of state of the circular string
solutions in de Sitter and anti de Sitter
spacetimes. The string equation of state has the perfect fluid form
$P=(\gamma-1)E,$ with $P$ and $E$ expressed closely and completely in
terms of elliptic functions (see Ref.\cite{all2} for the details)
and the instantaneous parameter $\gamma$ depending
on an elliptic modulus. We have quantized the time-periodic (oscillating)
string solutions within the semi-classical (stationary phase approximation)
approach and found the mass formulas in de Sitter and anti de Sitter
spacetimes. The
semi-classical quantization of the {\it exact} circular string solutions
and the canonical quantization of generic strings
(string perturbation series
approach), provide the same qualitative results.
\vskip 48pt
\hspace*{-6mm}{\bf Acknowledgements:}\\
A.L. Larsen is supported by the Danish Natural Science Research
Council under grant No. 11-1231-1SE

\newpage
\begin{centerline}
{\bf Figure Captions}
\end{centerline}
\vskip 24pt
\hspace*{-6mm}Figure 1. The potential $V(r^2)=r^2(1-H^2r^2),$ defined in
Eq.(2.3). For
$bH^2\alpha'^2<1/4,$
it acts effectively as a barrier. The horizon corresponds to
$H^2r^2=1.$
\vskip 12pt
\hspace*{-6mm}Figure 2. The hyperboloid time $q_-^0,$ in the
$bH^2\alpha'^2=1/4$ case, as
a function of $\tau.$
Notice that $\tau$ is a two-valued function of $q_-^0.$
\vskip 12pt
\hspace*{-6mm}Figure 3. (a) The hyperboloid time $q^0_-$ as a function of
$\tau$ in the
elliptic case $bH^2\alpha'^2<1/4.$ Each of the infinitely many branches
corresponds to one string. (b) The comoving time $T_-$ as a function of $\tau$
in the elliptic case $bH^2\alpha'^2<1/4.$ Notice that $T_-$ is not periodic.
\vskip 12pt
\hspace*{-6mm}Figure 4. The energy and pressure for a string
expanding from $r=0$ towards infinity (unstable string)
in de Sitter spacetime. The curves are drawn for the case $bH^2\alpha'^2=0.3.$
\vskip 12pt
\hspace*{-6mm}Figure 5. The energy and pressure for an
oscillating (stable)
string in de Sitter spacetime. The curves describe one period of
oscillation. The curves are drawn for the case $bH^2\alpha'^2=0.15.$
\vskip 12pt
\hspace*{-6mm}Figure 6. (a) Parametric plot of $H^2\alpha'W$ as a function of
$H^2m^2\alpha'^2$ for $k\in[0,\;1]$ in de
Sitter spacetime. There is only a finite number of states.
(b) Parametric plot of $H^2\alpha'W$ as a function of
$H^2m^2\alpha'^2$ in anti
de Sitter spacetime. There are infinitely many states.
\newpage
\begin{centerline}
{\bf Table Captions}
\end{centerline}
\vskip 24pt
\hspace*{-6mm}Table 1. Circular string evolution
in de Sitter spacetime. For each value of
$bH^2\alpha'^2,$ there exists two independent solutions $r_-$ and $r_+$:
\vskip 12pt
\hspace*{-6mm}Table 2. Circular string
energy and pressure in Minkowski, de Sitter and anti de
Sitter spacetimes.
\vskip 12pt
\hspace*{-6mm}Table 3. Semi-classical quantization of oscillating
circular strings in Minkowski, de Sitter and anti de
Sitter spacetimes.
\end{document}